# PERFORMANCE EVALUATION OF PEDESTRIAN NAVIGATION ALGORITHMS FOR CITY EVACUATION MODELING

F. Haghpanah[1], J. Mitrani-Reiser[2], and B.W. Schafer[3]

## ABSTRACT

Simulation is a powerful tool to study the behavior of physical, environmental, and social systems under different conditions. Evacuation simulation can be used to estimate the required time for people to exit a building or evacuate disaster exposed regions. While building evacuation simulation has seen significant study, city evacuation simulation is less developed. For evacuation simulations using Agent-Based Models, the characteristics of the underlying navigation algorithms are important in the overall efficiency of the simulation. In some disasters, e.g. earthquakes, evacuation takes place after the main event. This means evacuating and navigating in an environment with damaged and collapsed buildings and bridges and obstructed roads and paths. Furthermore, possible aftershocks or induced phenomena, such as landslide and liquefaction, can render a more dynamic situation for evacuees where the physical environment changes through time. Evacuees, modeled as agents, require a reliable algorithm for their navigation in these complex dynamic environments. A reliable navigation algorithm should be capable of handling obstacles with different physical properties and performing through dynamic environments. In this study, a framework is introduced to evaluate the relative performance of agent navigation algorithms. The main indices of this framework are Convergence, Optimality, Precision, and Efficiency (COPE). The COPE framework is applied on a set of robot navigation algorithms (the Bug Family) to assess their suitability to be used as pedestrian navigation algorithms.

---

[1]Graduate Research Assistant, Dept. of Civil Engineering, Johns Hopkins University, Baltimore, MD 21218, USA (email: haghpanah@jhu.edu)
[2]Assistant Professor, Dept. of Civil Engineering, Johns Hopkins University, Baltimore, MD 21218, USA
[3]Professor, Dept. of Civil Engineering, Johns Hopkins University, Baltimore, MD 21218, USA

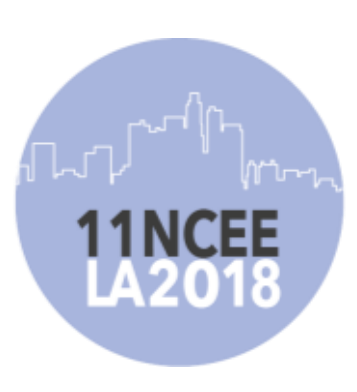

# Performance Evaluation of Pedestrian Navigation Algorithms for City Evacuation Modeling

F. Haghpanah[1], J. Mitrani-Reiser[2], and B.W. Schafer[3]

## ABSTRACT

Simulation is a powerful tool to study the behavior of physical, environmental, and social systems under different conditions. Evacuation simulation can be used to estimate the required time for people to exit a building or evacuate disaster exposed regions. While building evacuation simulation has seen significant study, city evacuation simulation is less developed. For evacuation simulations using Agent-Based Models, the characteristics of the underlying navigation algorithms are important in the overall efficiency of the simulation. In some disasters, e.g. earthquakes, evacuation takes place after the main event. This means evacuating and navigating in an environment with damaged and collapsed buildings and bridges and obstructed roads and paths. Furthermore, possible aftershocks or induced phenomena, such as landslide and liquefaction, can render a more dynamic situation for evacuees where the physical environment changes through time. Evacuees, modeled as agents, require a reliable algorithm for their navigation in these complex dynamic environments. A reliable navigation algorithm should be capable of handling obstacles with different physical properties and performing through dynamic environments. In this study, a framework is introduced to evaluate the relative performance of agent navigation algorithms. The main indices of this framework are Convergence, Optimality, Precision, and Efficiency (COPE). The COPE framework is applied on a set of robot navigation algorithms (the Bug Family) to assess their suitability to be used as pedestrian navigation algorithms.

## Introduction

In disaster studies, simulation is widely used to explore how natural hazards might evolve in the future, and how societies would react to these events. In many disasters, evacuation of buildings, neighborhoods, or urban regions is an important step towards ensuring public safety. Accordingly, evacuation simulation is a potentially helpful tool for emergency responders and policy makers to evaluate required time for evacuation and for estimating numbers and distribution of casualties under a disaster scenario.

[1]Graduate Research Assistant, Dept. of Civil Engineering, Johns Hopkins University, Baltimore, MD 21218, USA (email: haghpanah@jhu.edu)
[2]Assistant Professor, Dept. of Civil Engineering, Johns Hopkins University, Baltimore, MD 21218, USA
[3]Professor, Dept. of Civil Engineering, Johns Hopkins University, Baltimore, MD 21218, USA

Evacuation simulation can be classified into two main families: macroscopic models and microscopic models. Macroscopic models consider crowds as a whole, as in fluid-dynamic models, whereas microscopic models predict the crowd dynamics by considering individual behavior and interactions. Microscopic models can be discrete like Cellular Automata (CA) or continuous and dynamic, such as social force models and Agent-Based Models (ABMs) [1]. Each of these modeling approaches has specific advantages and disadvantages. Macroscopic models fail to incorporate social behavior of individuals in decision-making processes, and they are suitable only for environments where obstacles have rather simple shapes [2]. Although microscopic models can incorporate individual's behaviors, they are not free of deficiencies. CA models perform well for low to medium size crowds, but for highly crowded scenarios, the results could be unrealistic. The advantage of ABM is that by defining the behavior and rules on the microscopic level (i.e. the agents), diverse and unexpected macroscopic or mass responses can be observed [3]. Although ABMs are computationally more expensive than most other models, they possess the advantage of having the capability of implementing unique behaviors of heterogeneous individuals — an important feature which is not properly addressed in other models. ABMs are also difficult to implement due to complexities in defining logical rules for human behaviors and decision making processes; i.e. a need for sophisticated cognitive models for human behaviors. ABM is a bottom-up modeling approach where complicated global behavior of a system or process can be predicted by modeling the smallest elements (agents) of the system and defining their behavior through different situations. Furthermore, other types of models, such as social force rules or CA, can be incorporated into ABM. There have been numerous studies on evacuation simulation of buildings and urban regions using ABM [2, 4-9].

In general, pedestrian movements are different than vehicle movements. Vehicles have to follow lane boundaries and switch lanes when it is required, or if decided, given it is possible. However, pedestrian movements are subjected to more randomness such that each pedestrian has its unique trip toward the destination. There are different techniques to implement pedestrian movements, such as shortest path algorithms, potential field theory, and navigation algorithms. These techniques have relative advantages and disadvantages when the size of the model (i.e. number of agents, size of the simulation environment, and sizes and shapes of the obstacles) is large, as in city evacuation simulation. Those using shortest path algorithms, such as the Dijkstra's algorithm, although can generate locally optimal paths, they need preprocessing of the simulation environment to develop origin-destination paths for all nodes in the model. These models are not suitable for city evacuation modeling when accounting for damage conditions, where there could be obstacles with complex non-convex shapes and too many vertices. This affects model efficiency while the implementation is difficult, as well. The situation could be more complex if the model incorporates dynamic features of an event, such as flood propagation or possible collapse of buildings and road network during evacuation, where the origin-destination path generation process should be repeated at every time step. Those models using navigation algorithms do not need any preprocessing; however, the navigation algorithm must be reliable, efficient, and capable of handling obstacles of any shapes and sizes.

To identify reliable agent navigation algorithms for city evacuation modeling, a performance evaluation framework is developed. The framework consists of a list of typical obstacles that can be challenging for navigation algorithms to process (in the context of city evacuation), and a set of performance indices. A selection of algorithms from the Bug family is evaluated using this framework, and their relative performances are compared.

## COPE: A Performance Evaluation Framework

The COPE (Convergence, Optimality, Precision, and Efficiency) performance evaluation framework for pedestrian navigation algorithms consists of three main components. First, a list of benchmark obstacles that have been shown to be challenging based on the literature and modeling experience. Second, a performance evaluation metrics as elaborated in the following section. Third, the navigation algorithms to be evaluated. It should be emphasized that this framework provides a relative evaluation. Adding or removing algorithms will change the results.

### Benchmark Obstacles

For this study, five typical challenging obstacles that are potentially challenging for the navigation algorithms to process are identified (see Fig. 1): a long L-shaped obstacle, a U-shaped obstacle (with the target inside or outside of the obstacle), an obstacle with a pixelated edge, a T-shaped corridor, and a closed box obstacle.

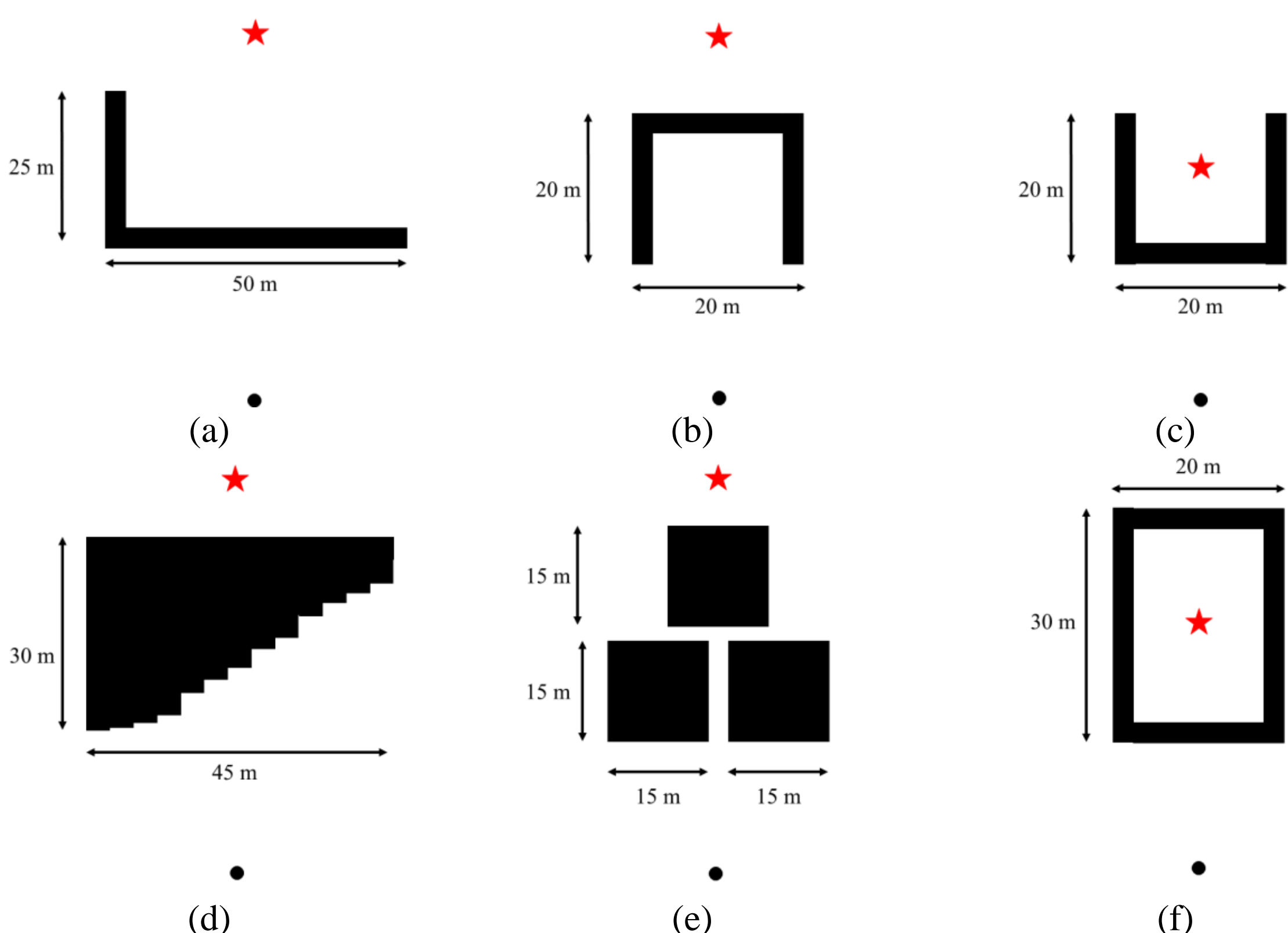

Figure 1. Benchmark obstacle: (a) long L-shaped obstacle, (b) U-shaped obstacle with target outside, (c) U-shaped obstacle with target inside, (d) obstacle with pixelated edge, (e) T-shaped corridor, and (f) closed box obstacle. Legend: black circle: agent; red star: target; black blocks: obstacle.

## COPE Performance Evaluation Metrics

The COPE performance evaluation metrics consider four indices taking values from 0 to 1: (1) Convergence, (2) Optimality in terms of length of generated path, (3) Precision in flagging trapped agents, and (4) Efficiency in terms of computation run-time. For each algorithm-obstacle pair, convergence is evaluated by the ratio of number of converged (i.e. successful) simulations to total number of simulations. Non-convergence is defined as cases where an agent is stuck in an infinite loop and cannot reach its destination. Optimality is evaluated by normalizing the length of the generated path with respect to the length of the optimal path considering all the algorithms. Regarding Precision, some algorithms define specific criteria to identify trapped agents, but these criteria may fail to or incorrectly flag agents as trapped. Precision in trap-flagging is evaluated by the ratio of number of false flagged (false positive) and false non-flagged (false negative) agents to total number of agents. Efficiency is evaluated by normalizing the run-time it takes for an algorithm to finish the simulation with respect to the fastest algorithm. The COPE indices for each algorithm-obstacle pair can be calculated using Eq. 2. The average values of COPE indices are then taken over the set of obstacles for each algorithm, as in Eq. 3. The total performance COPE Index for each algorithm can be obtained using Eq. 1 providing an index from 0 to 1. A higher COPE Index implies a relatively better performance.

$$I_i = \frac{w_c C_i + w_o O_i + w_p P_i + w_e E_i}{w_c + w_o + w_p + w_e} \tag{1}$$

$$C_{ij} = \frac{S_{ij}}{N_{ij}}, \quad O_{ij} = \frac{\min_i(L_{ij})}{L_{ij}}, \quad P_{ij} = 1 - \frac{F_{ij}}{N_{ij}}, \quad E_{ij} = \frac{\min_i(T_{ij})}{T_{ij}} \tag{2}$$

$$C_i = \sum_{j=1}^{n} C_{ij}, \quad O_i = \sum_{j=1}^{n} O_{ij}, \quad P_i = \sum_{j=1}^{n} P_{ij}, \quad E_i = \sum_{j=1}^{n} E_{ij}$$
$$i \in \{\text{set of algorithms}\} \quad , \quad j \in \{\text{set of obstacles}\} \tag{3}$$

where for algorithm $i$ and obstacle $j$, $S_{ij}$ is the total number of converged runs, $N_{ij}$ is the total number of runs, $L_{ij}$ is the length of the generated path, $F_{ij}$ is the total number of agents falsely flagged or not flagged as trapped, $T_{ij}$ is the computational time, $n$ is the number of obstacles, and $w_c$, $w_o$, $w_p$, and $w_e$ are weights for convergence, optimality, precision, and efficiency, respectively. For this study, $w_c = 5$, $w_o = 2$, $w_p = 1$, and $w_e = 2$.

## Navigation Algorithms

The Bug Algorithm Family is a family of robot navigation algorithms that gives logical solutions when no global information of the environment is available. These algorithms use range sensors and/or tactile sensors to identify obstacles and find a way to pass through and reach the destination. Some of these "Bug algorithms" can do better than others for a given environmental setting, but may perform weaker for other different settings [10]. Since evacuees take locally optimal paths when passing obstacles [11,12], algorithms that perform logically for the most different environmental settings have the highest potential for agent navigation in city evacuation simulation. The navigation algorithms considered for this study are: Bug1 and Bug2 [13], DistBug [14], KBug [15], and TangentBug [16]. DistBug comes with three optional extensions that can improve its performance. Among these algorithms, Bug1 and Bug2 identify obstacles when the agent hit one, TangentBug identifies obstacles using a radius of vision, and DistBug and KBug use a combination of both.

## Results

For each algorithm-obstacle pair, 1000 simulations are conducted to account for possible initial positions and local directions of agents. In each simulation, a target is placed on one side of the obstacle with an agent on the other side, as in Fig. 1. The simulations are implemented in NetLogo [17] using a computer with an Intel Zeon E3-1505M v5 @ 2.8 GHz processor and a 32GB memory. The results of the simulations are presented in Figs. 2 to 7 in the form of graphical traces for a selection of algorithms.

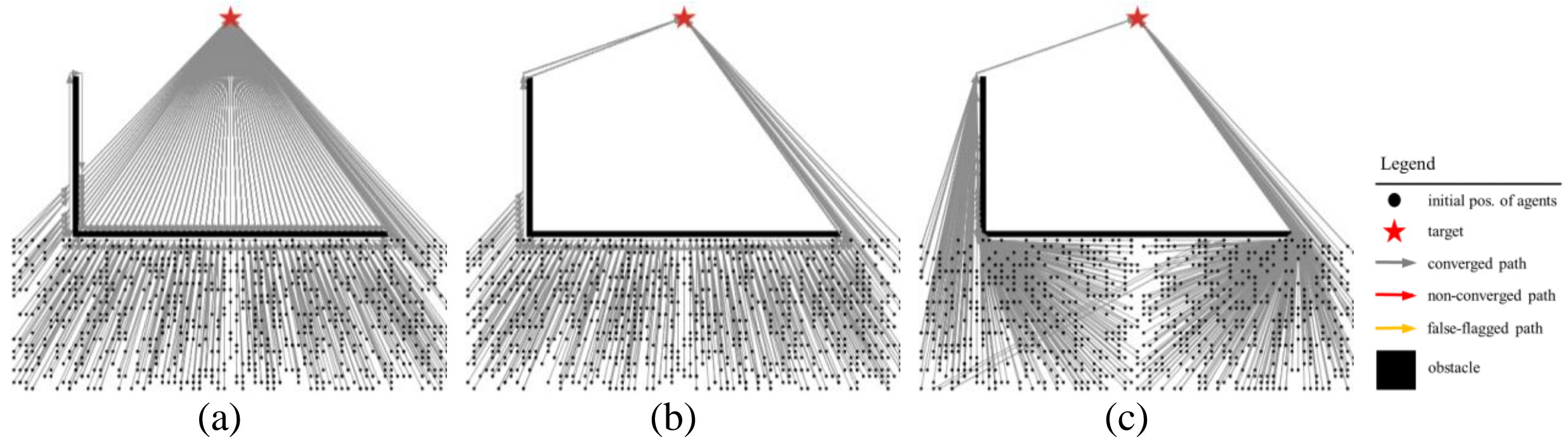

Figure 2. Results for the L-shaped obstacle: (a) Bug2, (b) DB1, and (c) KBug.

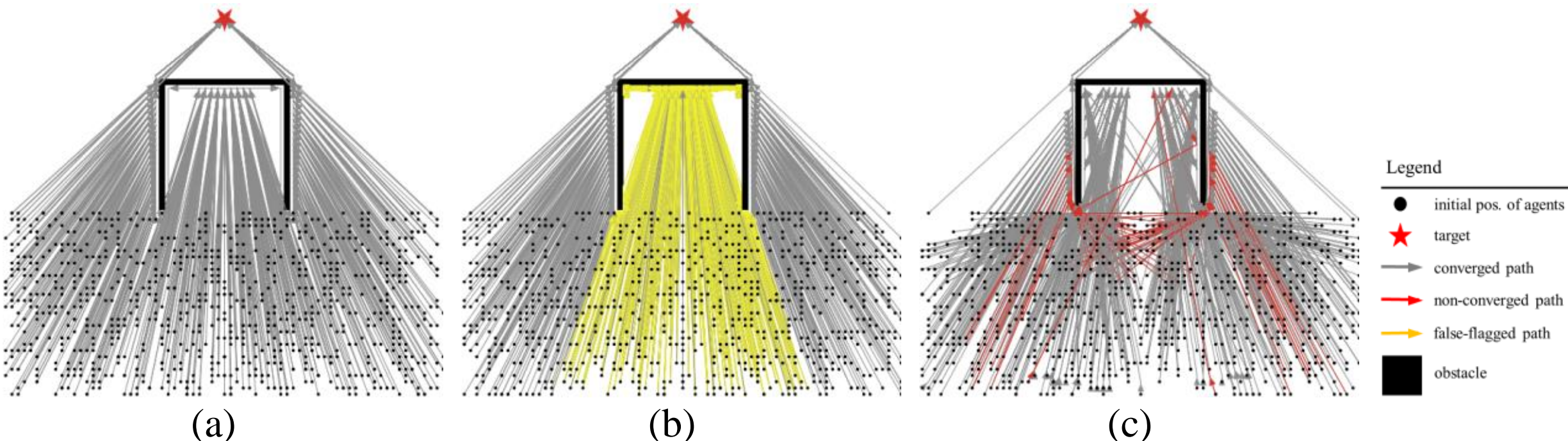

Figure 3. Results for the U-shaped obstacle (target outside): (a) DB1, (b) DB123, and (c) KBug.

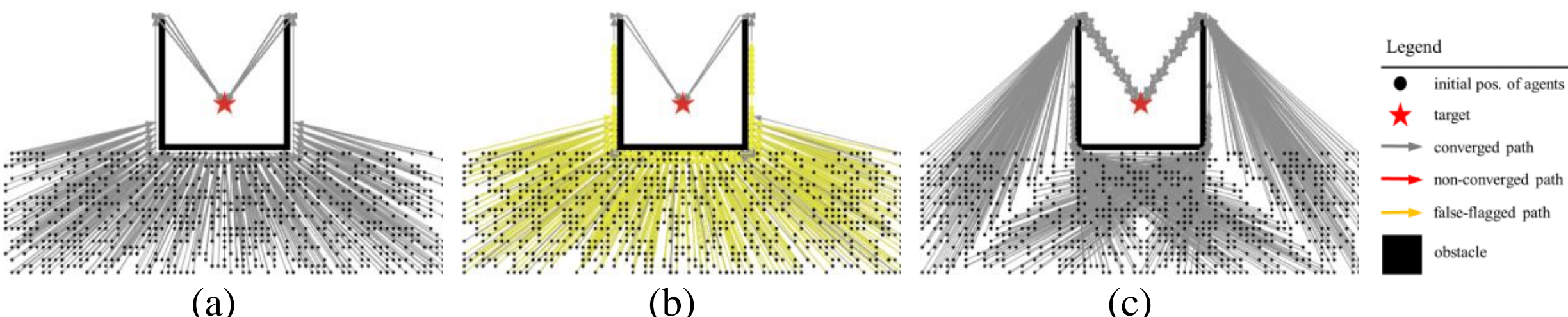

Figure 4. Results for the U-shaped obstacle (target inside): (a) DB1, (b) DB123, and (c) KBug.

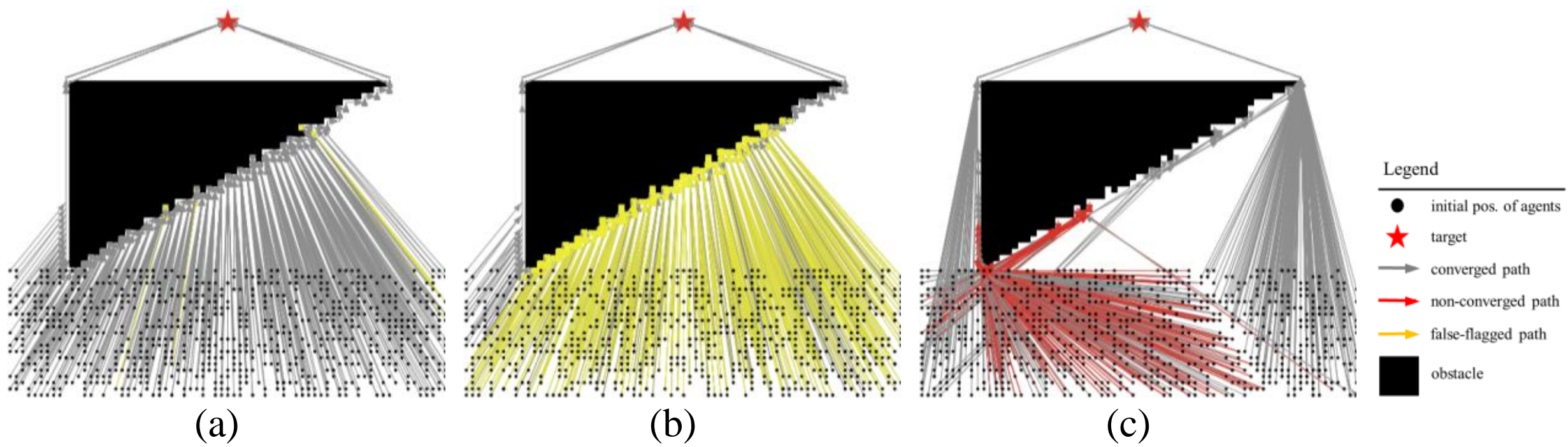

Figure 5. Results for the pixelated obstacle: (a) DB, (b) DB123, and (c) TangentBug.

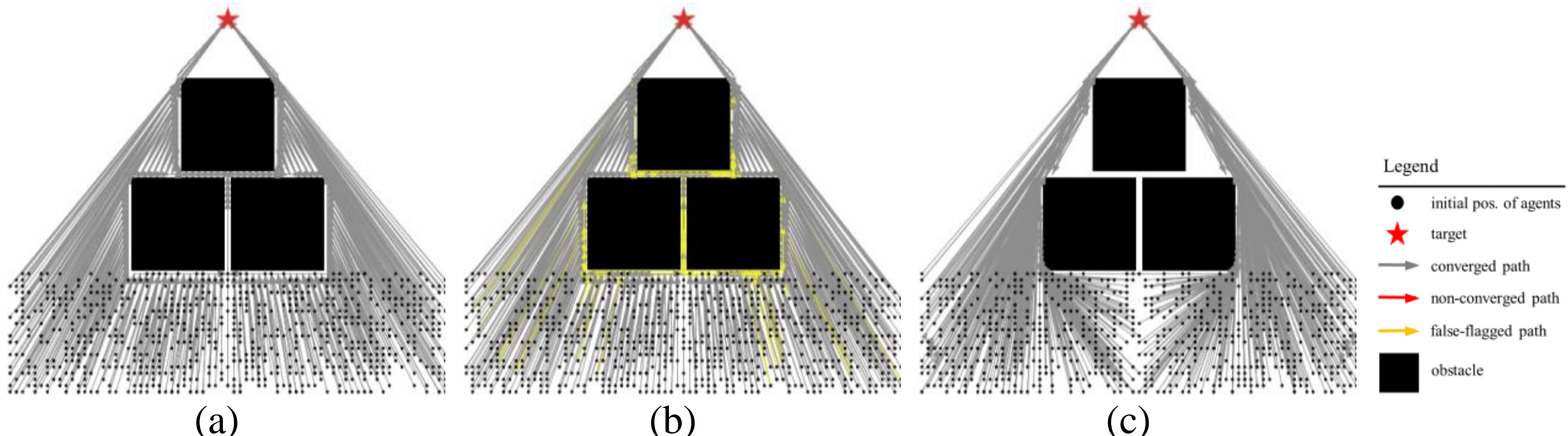

Figure 6. Results for the T-shaped corridor: (a) DB1, (b) DB3, and (c) KBug.

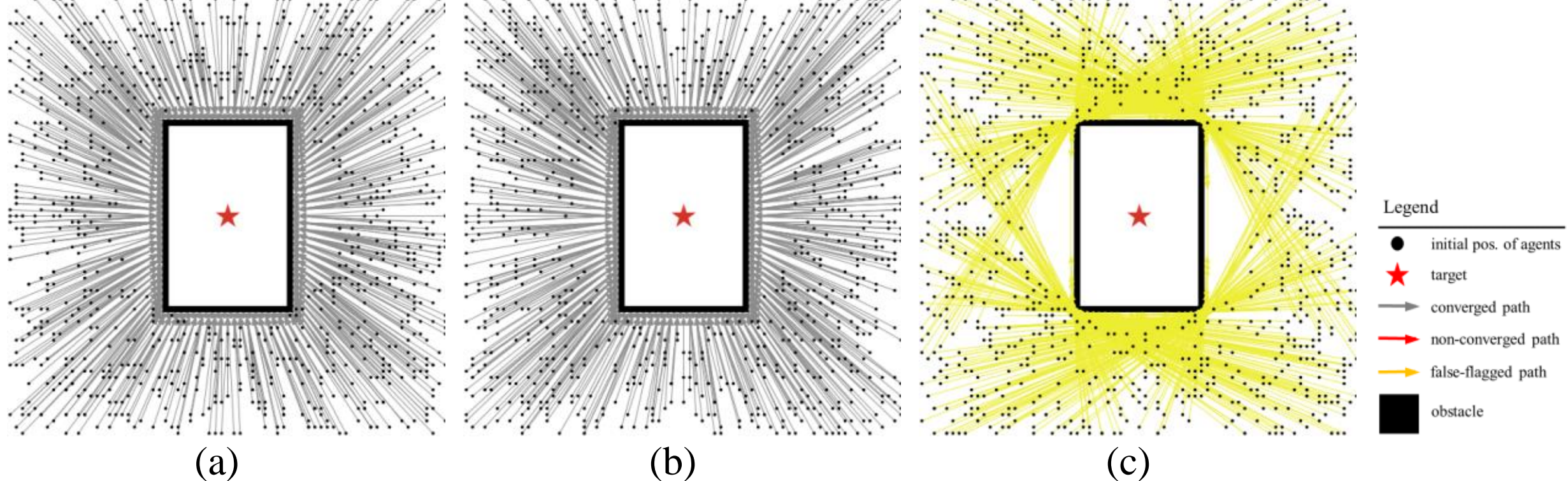

Figure 7. Results for the closed box obstacle: (a) Bug2, (b) DistBug, and (c) KBug.

## Discussion

All navigation algorithms are developed such that theoretically they can navigate any obstacle. In this section, we will explore the results of the simulations, and describe why some of the algorithms fail to perform for certain obstacles.

*Bug1 Algorithm.* According to the Bug1 algorithm, when an agent reaches the leave point on the boundary of an obstacle, if it identifies another obstacle in front, it will consider itself as trapped. In environmental settings such as the T-shaped corridor or wide obstacles with pixelated edges where the space between obstacles' corners is small (one patch in the context of this study), this leads to false-flagging agents as trapped. Fig. 8 (a) shows an example in which the red cross is where the agent is falsely flagged as trapped.

*Bug 2 Algorithm.* According to the Bug2 algorithm, when an agent is following the boundary of an obstacle, if it returns to the hit point, it considers itself as being trapped. This could be problematic where obstacles have pixelated edges (see Fig. 8 (b)). In general, aside from this error, the algorithm is simple, fast, works well for obstacles of any shapes, and generates shorter paths than Bug1, except for maze-like obstacles where the agent might get into cycles leading to longer paths [13].

*KBug Algorithm.* The KBug algorithm does not converge for obstacles with too many corners or pixelated edges. Agents get stuck along the corners and are not able to find a path to pass the obstacle (see Fig. 8 (c)). This is an acknowledged issue in the robot navigation literature; it is intuitive and recognized that obstacles with too many vertices are difficult for algorithms to process, particularly for those using a range sensor [18]. In general, KBug does not perform well when obstacles or obstacles' vertices are close to each other, it does not provide any criteria to identify trapped agents, and it is relatively slow because of constant screening of the environment; however, it can generate rather optimal (i.e. realistic) paths for different obstacles when there is no convergence issue.

*DistBug Algorithm.* DistBug is developed based upon Bug2 with improved leave conditions; therefore, it generates more optimal paths if the visual radius is large enough. However, since it is basically the Bug2 algorithm, it has Bug2's error in falsely identifying trapped agents. Moreover, the optional Extension-2 and Extension-3 lead to unnecessary change of direction when an agent is following the obstacle boundary, which causes the agent to be flagged as trapped when reaching a relatively wide obstacle with respect to its visual radius (see Fig. 8(d)). In addition, the generated paths are not optimal due to the change of direction while following the obstacle boundary with the extensions. These extensions to DistBug work well only if the size of the obstacle is relatively small with respect to the agent's visual radius. In general, excluding Extension-2 and Extension-3 rules, DistBug generates fairly optimal paths, and is relatively more efficient to compare with Bug1, Bug2, and KBug algorithms.

*TangentBug Algorithm.* TangentBug can generate solutions that approach the globally optimal paths when the environmental setting is simple and the agent's visual radius is relatively large with respect to the size of the obstacles [16]. Theoretically, the algorithm can handle obstacles of any shapes while generating optimal paths; however, it is not efficient for pedestrian navigation in which obstacles could be large with respect to the agent's visual radius; unlike in robot navigation, where obstacles are not wide relative to the robot's visual radius. This does not lead to failure of convergence, but it makes agents show unrealistic behavior (see Fig. 8 (e)). A solution is to set the visual radius of the agents to a large number, but this might be unrealistic depending on the scale of obstacles, health or age status of individuals, and the maximum visible distance (e.g. in case of evacuation involving fire and smoke), and it will make the algorithm more expensive. Another challenge in the implementation of TangentBug is building the Local Tangent Graph (LTG) and its nodes for obstacles with complex and non-convex shapes and for narrow pathways. Failure to build the perfect LTG leads to failure in finding the optimal path. It also leads to convergence issues and false-flagging agents as trapped. In theory, the LTG should be continuous, but in practice, the implementation of the LTG can be challenging. This can lead to error in identifying nodes [10]. The most prominent disadvantage of this algorithm is its computational cost. The algorithm takes much more time (of an order of 1000 to 10,000 in NetLogo) than other algorithms to finish a simulation. This is a huge drawback in the use of TangentBug for city evacuation simulation.

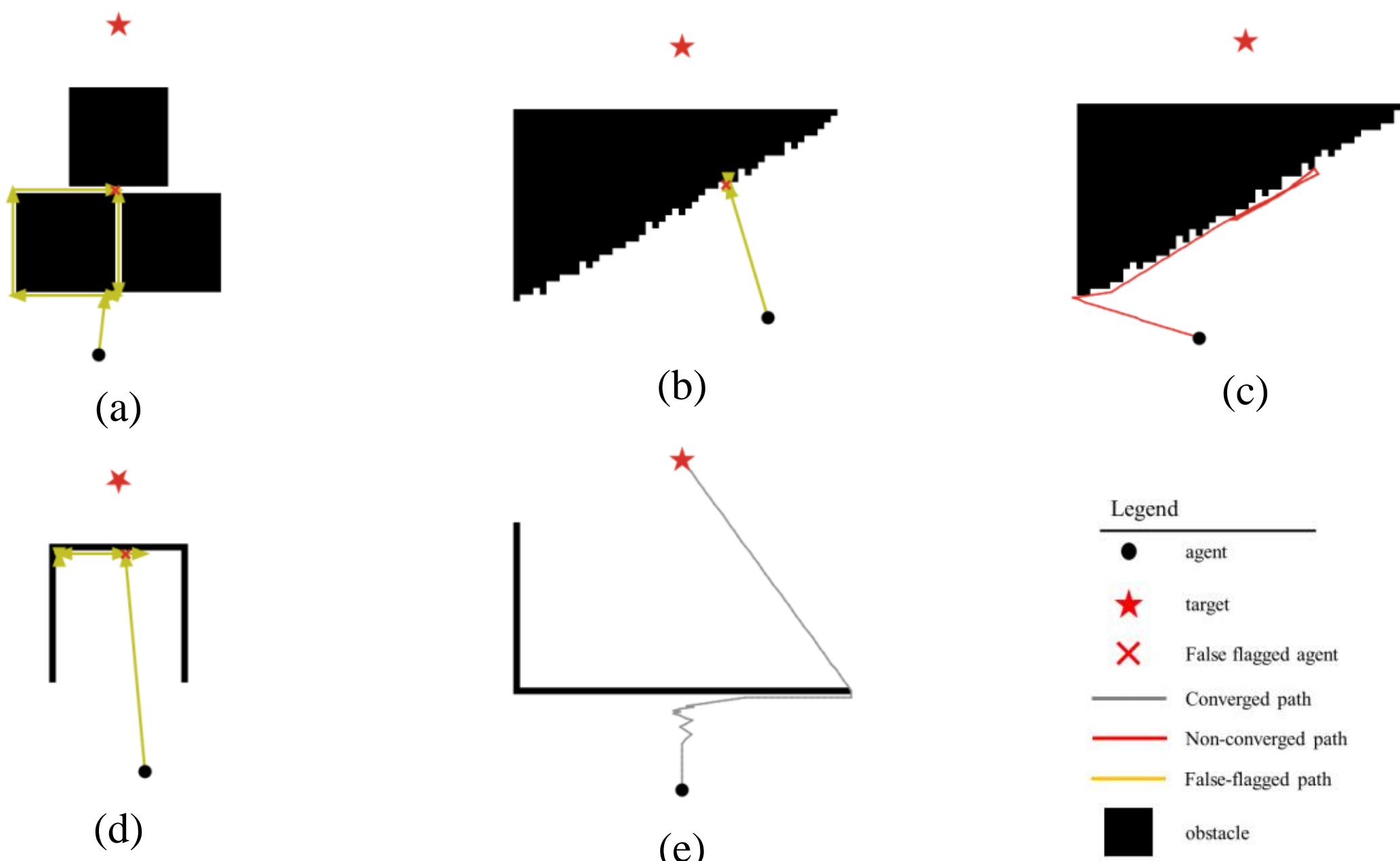

Figure 8. Error examples for: (a) Bug1; (b) Bug2 (c) KBug; (d) DistBug, and (e) TangentBug.

The results of the performance evaluation are presented in Figs. 9 and 10. In total, DistBug and its extensions have similar overall performances while DB1's performance is the highest with a COPE Index of 0.95 and therefore the best candidate among this selection of algorithms. DB1 converges for all the benchmark obstacles, provides rather optimal paths among the select algorithms, can identify trapped and non-trapped agents perfectly, except for the case of the pixelated obstacle where its Precision index is 0.97, and is the second most efficient algorithm (after DistBug) in terms of computation run-time.

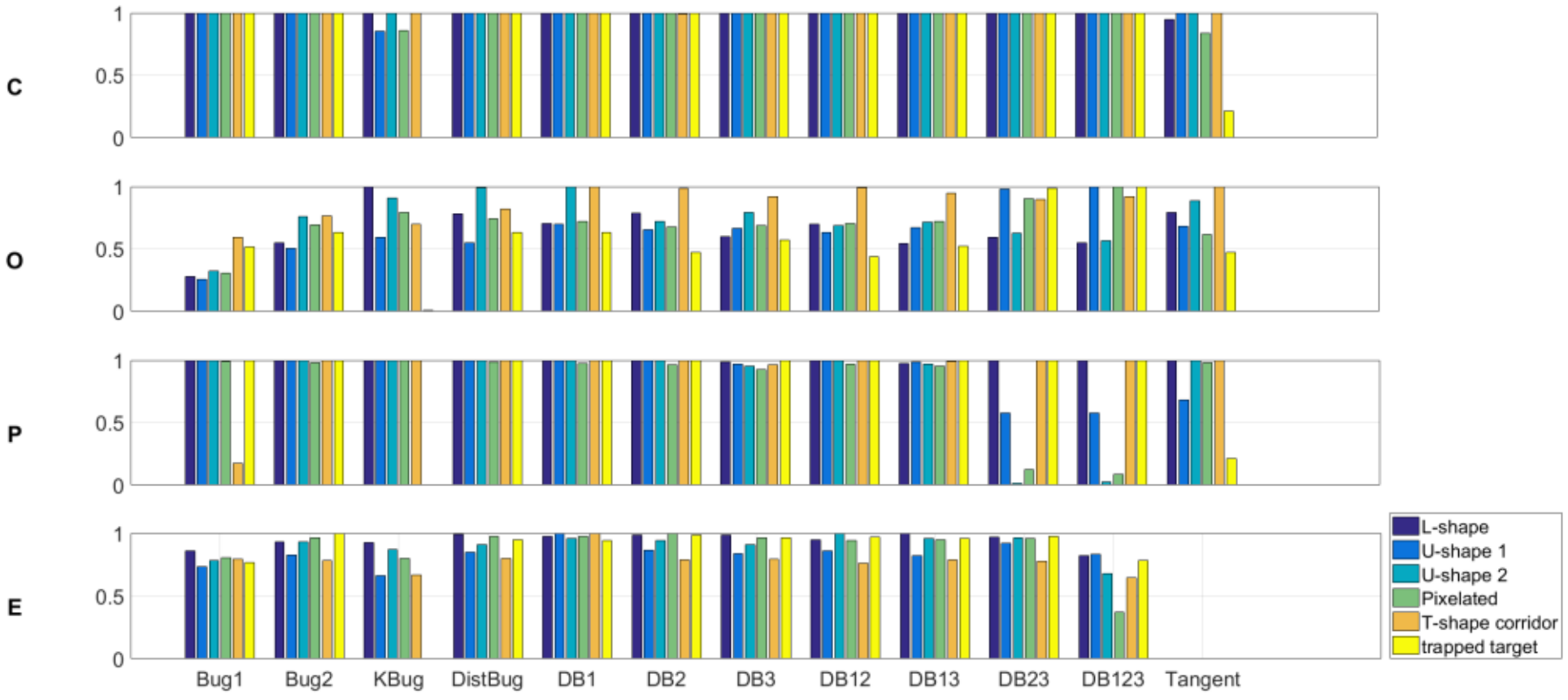

Figure 9. COPE indices grouped by algorithms

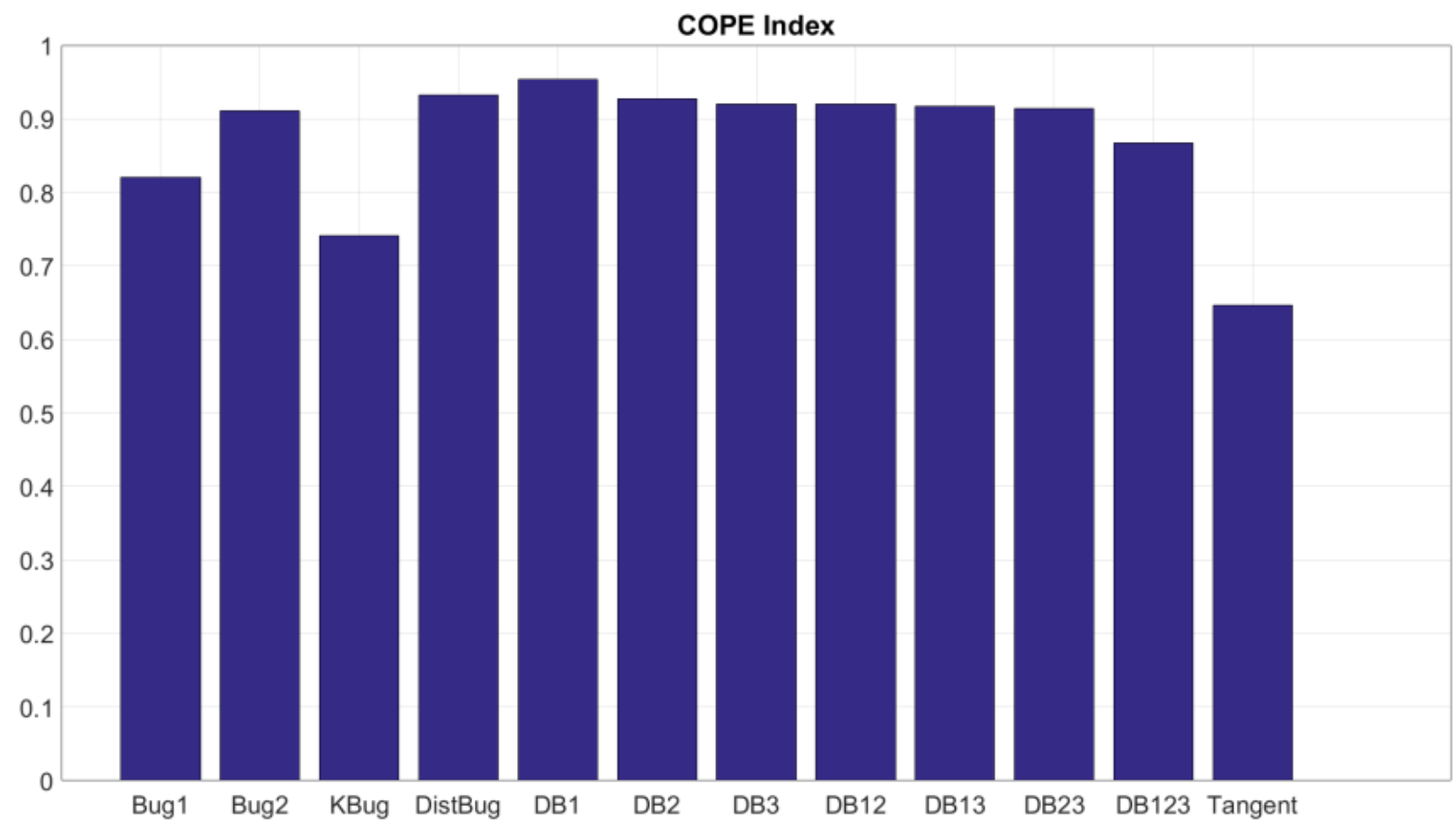

Figure 10. COPE index for the studied algorithms

## Conclusions

A competitive city evacuation model should be capable of accounting for damage conditions, where there could be obstacles with complex non-convex shapes. The environment could be more complex if the model considers dynamic features of an event, such as flood propagation, wildfires, or possible collapse of buildings and the road network during the evacuation. For such a model, a reliable pedestrian navigation algorithm is needed to be capable of processing obstacles of any shapes and sizes. To find reliable walking algorithms for city evacuation modeling, a performance evaluation framework is developed. The framework defines a list of typical obstacles that can be challenging for navigation. The performance of navigation algorithms, then, can be evaluated using the proposed performance indices. These indices are Convergence, path Optimality, Precision in flagging trapped people, and computational Efficiency (COPE). A selection of navigation algorithms from the Bug Family, namely Bug1, Bug2, KBug, DistBug, and TangentBug, is evaluated using these performance indices. Among these algorithms, TangentBug and DistBug perform relatively better in terms of generating optimal paths. However, accounting for other performance indices, DistBug has the relatively best performance. It is important to note than the results of this study are specific to modeling on the NetLogo platform where the physical space is represented in pixels. Other modeling platforms with different properties may result differently.

A future development of this study can be focused on the evaluation of more navigation algorithms to develop robust and reliable pedestrian navigation algorithms. Moreover, the performance of these navigation algorithms must be evaluated in real environmental setting. Furthermore, pedestrian behaviors such as queuing, crowd avoidance, and human-vehicle interactions could be included in the COPE performance evaluation framework.

## Acknowledgments

This work was funded through the National Science Foundation Grant RIPS 1441209. Its contents are solely the responsibility of the authors and do not necessarily represent the official views of the Centers for Disease Control and Prevention.